\documentclass{emulateapj}
\usepackage{apjfonts}
\usepackage{epsfig}

\begin{document}

{\small SLAC-PUB-12173}

\shorttitle{Gamma Rays from Dark Matter Satellites}
\shortauthors{Baltz, Taylor \& Wai}

\title{Can Astrophysical Gamma Ray Sources Mimic\\Dark Matter
Annihilation in Galactic Satellites?}

\author{Edward~A.~Baltz\altaffilmark{1}, James~E.~Taylor\altaffilmark{2} \&
  Lawrence~L.~Wai\altaffilmark{1}}

\altaffiltext{1}{KIPAC, Stanford Linear Accelerator Center, MS 29, 2575 Sand
  Hill Road, Menlo Park, CA 94025}

\altaffiltext{2}{Department of Physics and Astronomy, University of Waterloo,
  200 University Avenue West, Waterloo, Ontario, Canada N2L3G1}

\begin{abstract}
The nature of the cosmic dark matter is unknown.  The most compelling
hypothesis is that dark matter consists of weakly interacting massive particles
(WIMPs) in the 100 GeV mass range.  Such particles would annihilate in the
galactic halo, producing high-energy gamma rays which might be detectable in
gamma ray telescopes such as the GLAST satellite.  We investigate the ability
of GLAST to distinguish between WIMP annihilation sources and astrophysical
sources.  Focusing on the galactic satellite halos predicted by the cold dark
matter model, we find that the WIMP gamma-ray spectrum is nearly unique;
separation of the brightest WIMP sources from known source classes can be done
in a convincing way by including spectral and spatial information.  Candidate
WIMP sources can be further studied with Imaging Atmospheric Cerenkov
Telescopes.  Finally, Large Hadron Collider data might have a crucial impact on
the study of galactic dark matter.
\end{abstract}

\keywords{dark matter --- elementary particles --- Galaxy: halo --- gamma
rays: theory}

\section{Introduction}

It is now firmly established that the majority of matter in the universe is
non-baryonic.  Evidence for this standard cosmology includes the microwave
background anisotropies \citep{wmap} and the power spectrum of density
fluctuations on galactic scales \citep{sdss}.  The ``dark matter'' is of
unknown composition, but indirect evidence from particle physics and cosmology
indicates that it is likely to consist of weakly interacting massive particles
(WIMPs) in the mass range 30 GeV to 3 TeV.  Such particles would be expected to
annihilate slowly in galactic halos.  In most WIMP models, a large fraction of
the annihilation radiation is expected to be gamma rays from the decays of the
$\pi^0$ meson, produced copiously in any energetic interaction involving
hadrons.

In the cold dark matter (CDM) paradigm \citep{cdm1,cdm2}, it is well known that
structure forms hierarchically: the dark halos of galaxies such as the Milky
Way are expected to contain large numbers of sub-halos.  For WIMPs, the
sub-halo mass spectrum is expected to extend down to $10^{-6} M_\odot$
\citep{earthmass1,earthmass2}.  The substructure is expected to be nearly
isotropic, thus annihilation in the sub-halos can be away from the galactic
plane, where astrophysical sources are concentrated.

The brightest source of WIMP annihilation radiation is expected to be the
galactic center (GC), where the WIMPs are most concentrated.  The HESS
collaboration has concluded that dark matter annihilation radiation is at most
a small fraction of the emission coming from the GC above 100 GeV
\citep{hessgc}.  The MAGIC collaboration has confirmed these results
\citep{magicgc}.  To explain these data in terms of dark matter annihilation,
WIMPs of order 10 TeV mass would be required.  We will proceed with the
plausible assumption that the GC emission is predominantly astrophysical.  In
fact it has been shown that there is only a narrow window in which dark matter
annihilation could be discovered at the GC, given the HESS source and its
extrapolation below 100 GeV \citep{gcproblem}.  This emphasizes the point that
galactic satellites may be the most promising sources of WIMP annihilation
radiation.\footnote{There is a range of predictions for the annihilation rate
  at the GC.  Baryons are crucial: a massive bar can disrupt the central cusp,
  adiabatic contraction can strengthen it.  Satellites are simpler as their
  baryons can not cool.  The GC brightness is thus decoupled from the
  brightness of satellites.}

Many authors have discussed the possibility of annihilations in galactic
substructure \citep{BEGU,baltzdSph,CRM,Tasitsiomi,Stoehr,taylorsilk,
  evans,aloisio,Koushiappas,bi06,vialactea}; In this Letter, we will illustrate
that WIMP annihilation sources are distinguishable from all known (observed)
astrophysical source classes.  The detection of a steady, extended high
latitude source with a WIMP annihilation spectrum (e.g.\ by the GLAST
satellite) would provide strong evidence that the dark matter in the Galaxy
actually consists of particles in the 100 GeV mass range, a crucial piece of
the dark matter puzzle.

\section{Gamma Rays From Hadronic Interactions}
\label{sec:hadronicgamma}

Collisions of relativistic hadrons, such as when cosmic ray protons impinge on
the interstellar medium (ISM), are typically inelastic.  Energy is lost mostly
to $\pi^+,\pi^0,\pi^-$ mesons, in roughly equal numbers.  The decays of the
$\pi^0$ mesons dominate the gamma rays from hadronic interactions.  The $\pi^0$
has a mass $m_\pi=135.0$ MeV, and it has two common decay modes:
$\pi^0\rightarrow2\gamma$ (98.8\%) and $\pi^0\rightarrow e^+e^-\gamma$ (1.2\%)
with rare modes contributing less than 0.01\%.  As a pseudoscalar particle, it
decays isotropically.  The photons emitted in $\pi^0\rightarrow 2\gamma$ have
energies of $E_0=m_\pi/2=67.5$ MeV in the $\pi^0$ rest frame.  In the lab
frame, the energies are $E_\pm=E_0\gamma\left(1\pm\beta\cos\Theta_{\rm
  CM}\right)$.  The isotropy of the decay implies that $\cos\Theta_{\rm CM}$ is
uniformly distributed, and thus the spectrum $dN/dE$ is constant between the
minimum and maximum energies $E_{\rm min,max}=E_0\gamma\left(1\pm\beta\right)$.
The spectrum $dN/dE(\ln E)$ is symmetric about $\ln E_0$, because $E_{\rm
  max}/E_0=E_0/E_{\rm min}$.  The observed spectrum from a source will have
this property if the pion distribution is isotropic, true for both WIMP
annihilation and cosmic ray interactions.

\begin{figure}[h]
\epsfig{file=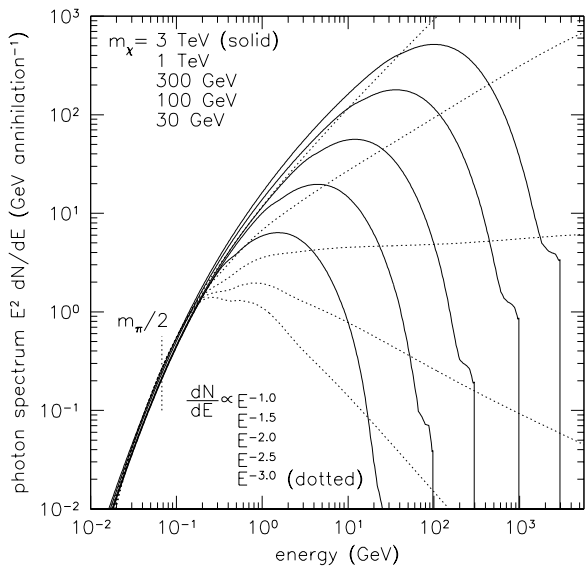, width=0.45\textwidth}
\caption{Spectrum of photons from hadronic processes.  Solid lines depict
annihilations to $b$ quark pairs for several WIMP masses.  The peak in the
spectrum occurs at an energy of $E_{\rm peak}\approx m_\chi/25$.  At low photon
energies the spectrum is nearly independent of WIMP mass in both shape and
magnitude.  Dotted lines depict $pp$ interactions for several proton spectra.
The $pp$ photon spectra are normalized to be equal at an energy of $E_0=67.5$
MeV (vertical dotted line), where their spectral slopes are guaranteed to be
equal.}
\label{fig:fullspectrum}
\end{figure}

The photon spectrum from a single pion energy must be convolved with the pion
spectrum.  We consider the process $\chi\chi\rightarrow b\overline{b}$, the
annihilations of pairs of self-conjugate dark matter particles to pairs of $b$
quarks.  The process is non-relativistic, giving monochromatic quarks with
energy $E_q=m_\chi$.  The quarks each form ``jets'' dominated by $\pi$ mesons.
In figure~\ref{fig:fullspectrum} we plot the photon spectra from annihilations
as calculated by DarkSUSY \citep{darksusy}, which uses results from Pythia
\citep{pythia}.  The spectrum is universal: even $\chi\chi\rightarrow
W^+W^-\;{\rm or}\;Z^0Z^0$ gives similar results.  Only the
$\chi\chi\rightarrow\tau^+\tau^-$ channel differs appreciably
\citep{cesarini,tautau,hoopertaylor}, but this is difficult to arrange for
WIMPs from supersymmetry.

In figure~\ref{fig:fullspectrum} we show the photon spectra from several
power-law proton sources as calculated by GALPROP \citep{galprop}.  This
illustrates that the power-law proton beams typical of astrophysical sources
can {\em not} easily mimic the gamma ray spectrum from WIMP annihilations.

\section{Detectability of Galactic Dark Matter Satellites}
\label{sec:satellites}

The GLAST satellite \citep{glast1,glast2,glast3} is well suited to measuring
gamma rays from dark matter annihilations.  It has an effective area of
$\approx$1 m$^2$, a solid angle acceptance of $\approx 3$ sr, and a
point spread function (PSF) of 0.4$^\circ$ at 1 GeV energy.  It will measure
gamma ray energies between 20 MeV and 300 GeV.  Most data will be taken in
survey mode, mapping the sky with equal coverage with a large duty cycle.  The
exposure to any source will reach roughly $3\times 10^{11}$ cm$^2$ s in a 5
year mission.

We have estimated the number of Milky Way dark matter satellites observable by
GLAST.  The calculation was performed with the semi-analytic method
of~\citet{taylor1,taylor2,taylor3}.  The satellite mass distribution has the
expected $dN/dM\propto M^{-2}$ \citep{clumpspectrum}, cutting off below $10^6
M_\odot$ due to computational limitations.  The spatial distribution of
satellites is roughly spherically symmetric about the galactic center and
extends well beyond the solar orbit, thus the dark matter satellites are
located mostly out of the galactic plane.  Individual sources have NFW density
profiles \citep{nfw}, with central $r^{-1}$ cusps.  Satellites with steeper
profiles, e.g.\ \citep{moore}, would be easier to detect.  We find that the
brightest sources have masses in the $10^6-10^7 M_\odot$ range.  These
brightest sources have tidal radii of order 100 pc, typically corresponding to
$1^\circ$ on the sky.  We note that most of these objects are severely
stripped.  They have scale radii $r_s$ much larger than their tidal truncation
radii $r_t$, thus they have nearly pure $r^{-1}$ density profiles out to
$r=r_t$.

The surface brightness in gamma rays is proportional to the parameter
$J\propto\int\rho^2\,dr$ \citep{BUB}.  For a stripped NFW clump, at a fixed
angular distance from its center, $J\propto M^2/r_t^4/D$, where $D$ is the
distance.  If the mass spectrum of clumps is $dN/d\ln\,M\propto M^{-\alpha}$
and the tidal radius $r_t\propto M^\beta$, the surface brightness of the
nearest clump ($D\propto M^{\alpha/3}$) is $J\propto M^{2-4\beta-\alpha/3}$.
Our simulations indicate that $\alpha\approx 1$ and $\beta\approx 1/2$, thus
$J\propto M^{-1/3}$.  Lower mass clumps are brighter, but they appear smaller
($\theta\propto r_t/D\propto M^{\beta-\alpha/3}\propto M^{1/6}$).  Less massive
clumps would be seen as point sources.  These results are sensitive to $\alpha$
and $\beta$, thus extrapolations are difficult.

\begin{figure}[h]
\epsfig{file=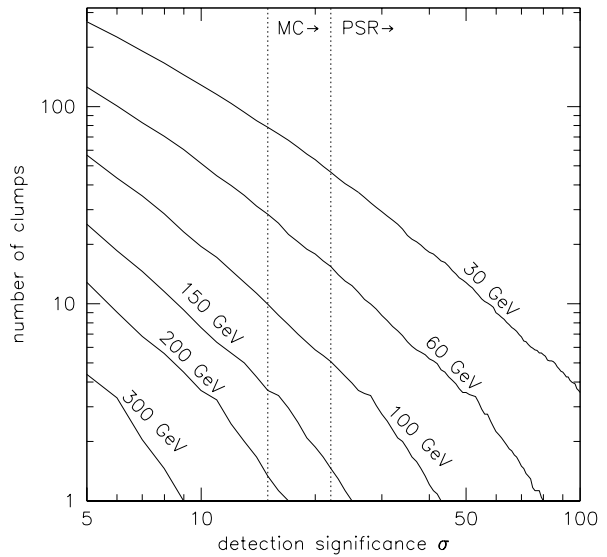, width=0.45\textwidth}
\caption{The number of detectable clumps for several WIMP masses is plotted
  against the detection threshold for counts above 1 GeV.  Distinguishing the
  WIMP spectrum at $3\sigma$ from a molecular cloud and a pulsar is possible at
  detection significances of 15$\sigma$ (290 counts, 375 background) and
  22$\sigma$ (425 counts) respectively (vertical dotted lines).}
\label{fig:nclumps}
\end{figure}

As a fiducial case, we assume a WIMP mass of 100 GeV and an annihilation
cross-section to $b\overline{b}$ of $\langle\sigma v\rangle= 1.6\times10^{-26}$
cm$^3$ s$^{-1}$, giving 14.2 photons per annihilation above 1 GeV.  Assuming a
5 year GLAST mission, and integrating a $1^\circ$ radius around the source, the
number of background counts is 375 (based on the EGRET extragalactic background
\citep{egret}).  The typical brightest clump has $\langle
J\rangle(1^\circ\;{\rm radius})=1400$.  For an example of such an object take
$2\times10^6\,M_\odot$, 3 kpc distant, tidal radius 50 pc, thus subtending
$1^\circ$ on the sky.  The number of signal counts above 1 GeV within $1^\circ$
for 30, 60, 100, 150, 200, 300 GeV WIMPs is 3450, 1680, 900, 520, 345, 190
respectively.  The number of detectable dark matter satellites is shown in
figure~\ref{fig:nclumps}.

\section{Astrophysical Sources}
\label{sec:sources}

A pure dark matter galactic satellite has four distinguishing characteristics:
hadronic spectrum from monochromatic quarks, spatial extent, lack of
variability, and no emission at other wavelengths except for very diffuse
inverse Compton and synchrotron \citep{diffuseIC}.  We will focus on the energy
spectrum, but we note that a satellite with an NFW profile has a brightness
proportional to $1/r$, meaning equal flux in equal width annuli.  With the
$0.4^\circ$ PSF of GLAST above 1 GeV, and the typical $1^\circ$ size, we expect
that the spatial extent should be detectable.  Naively, emission between
$r=0.5^\circ$ and $r=1^\circ$ has $58\%$ of the total significance.

To mimic a galactic satellite, a source would need to have a broken power law
spectrum, no counterparts in other wavelengths, and degree-scale extended
emission constant in time.  Each of these is exhibited by known sources, but
none exhibits them all.  A class of such sources would need to have identical
spectra (as measured), as dark matter satellites would.

\begin{figure}
\epsfig{file=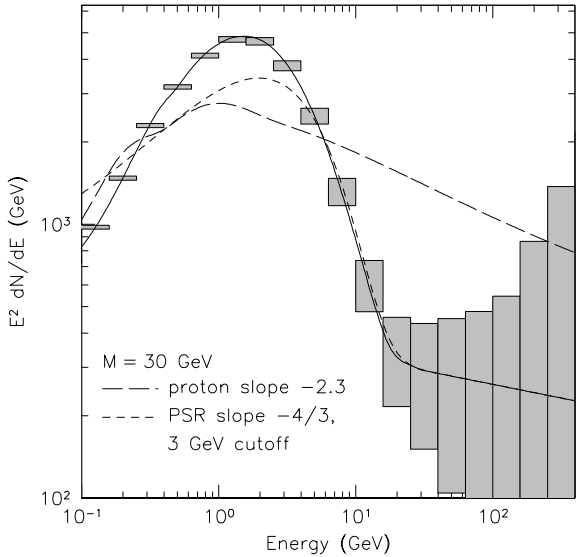, width=0.45\textwidth}
\caption{Annihilation of 30 GeV WIMPs.  The solid line represents the WIMP
  spectrum from a single clump with $\langle J \rangle=1400$ within $1^\circ$,
  including the extragalactic background.  Shaded boxes illustrate the
  $1\sigma$ (poisson) flux errors for 0.2 decade bins (the actual resolution is
  better than 10\%, and we expect very little dispersion between bins this
  large).  The best fit proton power law (long dashes) and \mbox{slope~-4/3}
  pulsar spectra (short dashes) are indicated.  These fits are unacceptable,
  including a pure power law (not shown).  This source would have had $\sim$50
  counts above 100 MeV in the third EGRET catalog, just above the detection
  limit \citep{3EG}.}
\label{fig:mockdata30}
\end{figure}

\begin{figure}
\epsfig{file=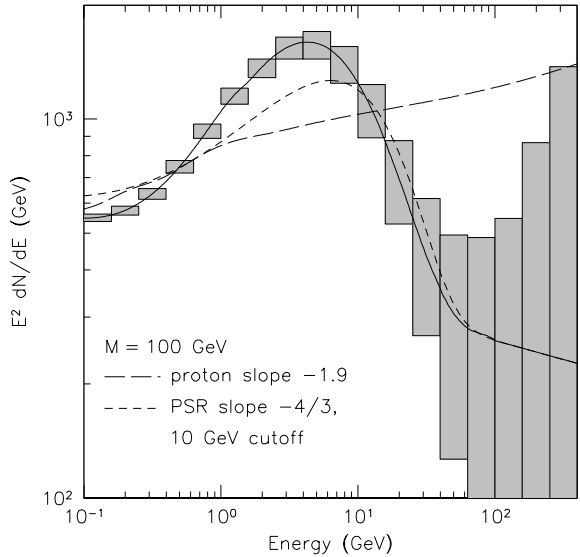, width=0.45\textwidth}
\caption{Annihilation of 100 GeV WIMPs.  The curves are the same as in
figure~\ref{fig:mockdata30}.  Again, none of the fits are acceptable, including
a pure power law.}
\label{fig:mockdata100}
\end{figure}

\begin{figure}
\epsfig{file=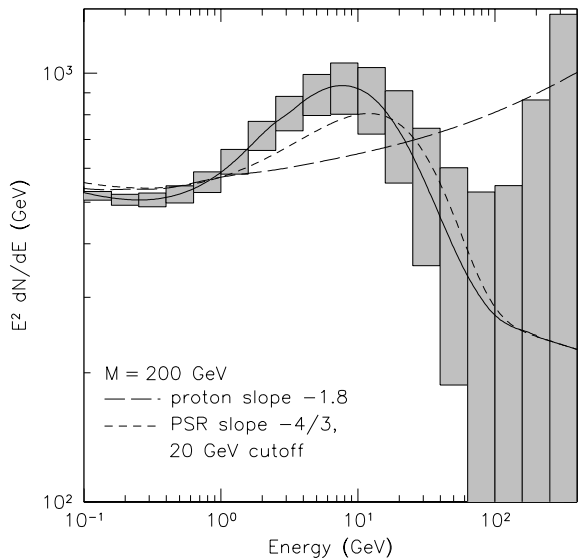, width=0.45\textwidth}
\caption{Annihilation of 200 GeV WIMPs.  The curves are the same as in
figure~\ref{fig:mockdata30}.  In this case, the pulsar fit is allowed at the
8\% level (by simple $\chi^2$), but the others are unacceptable, including a
pure power law.}
\label{fig:mockdata200}
\end{figure}

In figures~\ref{fig:mockdata30}-\ref{fig:mockdata200} we plot the spectrum of
the typical brightest clump $(\langle J\rangle=1400)$ with 30, 100, and 200 GeV
WIMPs together with fits for several astrophysical source classes.  The
$1\sigma$ errors for GLAST are shown as shaded boxes.  In
figure~\ref{fig:nclumps} we illustrate the detection significance required to
distinguish dark matter from molecular clouds and pulsars.

The gamma ray spectrum from molecular clouds is generated by cosmic ray
protons, which have a featureless power-law spectrum over many decades in
energy.  This is are exactly what is plotted with dotted lines in
figure~\ref{fig:fullspectrum}.  The long dashed lines in
figures~\ref{fig:mockdata30}-\ref{fig:mockdata200} show the best fit molecular
cloud spectra, ruled out at high confidence in each case.  Gamma rays from
molecular clouds are expected to be extended and non-variable; it would be
comforting to rule out counterparts, e.g. CO emission.  In fact, GLAST is
likely to detect extended, high-latitude molecular clouds \citep{molecular}.

Gamma ray pulsars are potentially the most problematic of the astrophysical
sources.  Their spectra can be parameterized as $dN/dE\propto
E^{-\Gamma}\exp[{-(E/E_c)^\alpha}]$~\citep{nel}.  The few known examples have
$\Gamma>4/3$.  In fact, most models for gamma ray pulsars require this
(e.g. the outer gap model \citep{outergap}), but $\Gamma\rightarrow 2/3$ is in
principle possible.

The short dashed lines in figures~\ref{fig:mockdata30}-\ref{fig:mockdata200}
show the best fit gamma ray pulsar spectra.  The $(\Gamma=4/3)$ spectrum is
ruled out except in the 200 GeV case, where it is consistent at the 8\% level.
If the low energy slope $\Gamma\rightarrow1$, the spectra become nearly
impossible to disentangle for any WIMP mass.

Gamma ray pulsars tend to have multi-wavelength counterparts and also tend to
be near the galactic plane.  A notable exception is 3EG J1835+5918 which is
located at high latitude, but does have a faint X-ray
counterpart~\citep{halpern}.  The well known radio quiet gamma ray pulsar
Geminga is located within $5^\circ$ of the galactic plane.

The variability of the pulsar is difficult to determine in a blind search of
the period-period derivative plane.  To mimic an extended galactic satellite, a
cluster of pulsars would be required, with no counterparts in other
wavelengths.

Plerions will typically have multi-wavelength counterparts, especially in
X-rays, and are located close to the galactic plane.  They are compact sources
in X-rays ($\sim 1'$), but at GLAST energies they may be detected as extended
sources.

Supernova remnants will have power-law spectra, convincingly ruled out.  They
will also have multi-wavelength counterparts and are likely to be near the
galactic plane.

Blazars are variable point sources with a power-law gamma ray spectrum, and
have counterparts.  They contradict all four necessary qualities of dark matter
satellite emission.

\section{Discussion}
\label{sec:discuss}

We have shown that the brightest dark matter satellites seen by GLAST should be
distinguishable from other known types of astrophysical sources, for WIMPs less
massive than about 150 GeV and cross sections of $\langle\sigma
v\rangle=1.6\times10^{-26}$ cm$^3$ s$^{-1}$.  Sub-substructure could enhance
these signals by a factor of a few \citep{vialactea}, extending the accessible
masses and cross sections.  Dimmer sources (extended only, to minimize pulsar
contamination) could be stacked, improving the discrimination.  Any sources
mimicking dark matter annihilation would be very interesting, and would be
compelling targets for study in a multiwavelength campaign.

Dark matter sources are excellent targets for Imaging Atmospheric Cerenkov
Telescopes (IACTs).  The mass of the WIMP must be above the IACT analysis
threshold, at the present time around 100 GeV.  The sensitivity of IACTs is
currently limited by the charged particle background.

A follow-up campaign of 500 hours with an IACT of 0.2 km$^2$ on our brightest
clump, taking the 100 GeV model, can provide a 5$\sigma$ detection of line
emission from the process $\chi\chi\rightarrow\gamma\gamma$, for a branching
ratio of $B=1.2\%$.  This assumes 99\% rejection of hadrons and 15\% energy
resolution.  If the hadron rejection were improved by a factor of 10, the
electron background dominates and line sensitivity improves to $B=0.005$.  If
the electron background were also eliminated, the extragalactic gamma-ray
background would limit the sensitivity to $B=0.0003$.  The predicted branching
ratio is $B\sim 0.001$.  Obviously, the detection of a line at these energies
would demonstrate the existence of particle dark matter.

The Large Hadron Collider (LHC) may discover a candidate WIMP, and measure its
mass at the 10\% level on a timescale that matches the GLAST program.  A simple
estimate shows that GLAST can constrain the mass at the 25\% level, for a 100
GeV WIMP.  If the GLAST and LHC mass estimates match, the WIMP hypothesis would
be greatly strengthened.  With strong evidence for particle dark matter in
hand, including accelerator measurements of cross sections \citep{BBPW}, it
would become possible to map the galactic dark matter in the gamma ray sky.

\acknowledgments

We thank Roger Blandford, Elliott Bloom, Jim Chiang, Stephan Fegan, Stefan
Funk, Joni Granot, Tune Kamae, Michael Peskin, Roger Romani, Marek Sikora and
Ping Wang for comments.
This work was supported in part by the Department of Energy contract
DE--AC02--76SF00515.

\end{document}